\documentclass[runningheads]{article}

\usepackage[T1]{fontenc}
\usepackage{graphicx}
\usepackage{authblk}
\usepackage{fancyhdr}
\newcommand{\titlerunning}[1]{\newcommand{\mytitlerunning}{#1}}
\newcommand{\authorrunning}[1]{\newcommand{\myauthorrunning}{#1}}
\newcommand{\institute}[1]{\affil{#1}}
\newcommand{\inst}[1]{$^{#1}$}
\newcommand{\keywords}[1]{%
  \par\noindent\textbf{Keywords:} #1
}
\newcommand{\ackname}{\subsection*{Acknowledgments}}
\usepackage{xcolor}
\usepackage{amsmath}
\usepackage[caption=false]{subfig}

\usepackage{hyperref}
\usepackage{cleveref} %
\usepackage{sidenotes} %
\usepackage[inline]{enumitem}
\setlist[enumerate]{leftmargin=*}
\usepackage{booktabs}
\usepackage{tikz}
\usepackage{pgfplots}
\pgfplotsset{width=10cm,compat=1.9}
\usepackage{pgfplotstable}

\tikzset{white border/.style={preaction={draw,white,line width=4pt}}}

\usepackage{listings}
\definecolor{BrickRed}{rgb}{0.8, 0.25, 0.33}

\newcommand{\myline}[1]{on line \texttt{\#}\texttt{#1}}
\newcommand{\mylines}[2]{on lines \texttt{\#}\texttt{#1-#2}}
\newcommand{\mylinestwo}[2]{on lines \texttt{\#}\texttt{#1} and \texttt{\#}\texttt{#2}}

\newcommand{\lstinprism}[1]{\lstinline[language=PRISM,columns=fullflexible,breaklines=true,basicstyle=\normalsize\ttfamily,postbreak=]{#1}}

\lstdefinelanguage{PRISM}{
  morekeywords={
      dtmc, mdp, ctmc, const, int, float, bool, global, module, init, endmodule, true, false, double,
      rewards, endrewards, formula, label
  },
  sensitive=true, %
  morecomment=[l]{//}, %
  morecomment=[is]{/*}{*/}, %
  morestring=[b]" %
} %

\definecolor{eclipseBlue}{RGB}{42,0.0,255}
\definecolor{eclipseGreen}{RGB}{63,127,95}
\definecolor{eclipsePurple}{RGB}{127,0,85}

\usepackage[colorinlistoftodos,textsize=tiny]{todonotes}
\usepackage{marginnote}

\hypersetup{
	colorlinks,
	linkcolor={blue!50!black},
	citecolor={blue!50!black},
	urlcolor={blue!80!black}
}

\usepackage[skins]{tcolorbox}
\usepackage[subfigure]{tocloft}
\usepackage{fmtcount}
\usepackage{tabularx,ragged2e}
\usepackage{makecell}

\usepackage{imakeidx}
\makeindex[title=Index of Changes, name=changes]

\newcommand{\notinpaper}[1]{%
	\index[changes]{Comment \textbf{C\expandafter\sortgref#1\sortgref} not reflected in the text.\removecomma|HIDE}%
}

\newrobustcmd{\removecomma}[1]{}
\newcommand{\HIDE}[1]{}

\newcommand{\C}[1]{%
	\index[changes]{Comment \textbf{C\expandafter\sortgref#1\sortgref}|BH{\arabic{changes}}}{C#1}%
}%

\newcommand{\Cn}[2]{%
  \index[changes]{Comment \textbf{C#1.\expandafter\sortgref#2\sortgref}|BH{\arabic{changes}}}{C#1.#2}%
}%

\def\sortgref#1\sortgref{%
	\ignoresort{\ifnum#1<10 00\else\ifnum #1<100 0\fi\fi#1}#1%
}
\protected\def\ignoresort#1{}

\begin{document}
\title{{Towards Achieving Energy Efficiency and Service Availability in O-RAN via Formal Verification}}
\titlerunning{Formal Verification for 6G O-RAN}
\author{%
Roberto Metere\inst{1}%
,
Kangfeng Ye\inst{1}%
,
Yue Gu\inst{2}%
,
Zhi Zhang\inst{3}%
,
Dalal Alrajeh\inst{3}
,
Michele Sevegnani\inst{2}%
,
Poonam Yadav\inst{1}
}
\authorrunning{R. Metere et al.}
\institute{%
\inst{1}University of York, UK
\\
\inst{2}University of Glasgow, UK
\\
\inst{3}Imperial College London, UK
}
\date{}
\maketitle              %
\begin{abstract}
As Open Radio Access Networks (O-RAN) continue to expand, AI-driven applications (xApps) are increasingly being deployed enhance network management.
However, developing xApps without formal verification risks introducing logical inconsistencies, particularly in balancing energy efficiency and service availability.
In this paper, we argue that prior to their development, the formal analysis of xApp models should be a critical early step in the O-RAN design process.
Using the PRISM model checker, we demonstrate how our results provide realistic insights into the thresholds between energy efficiency and service availability.
While our models are simplified, the findings highlight how AI-informed decisions can enable more effective cell-switching policies.
We position formal verification as an essential practice for future xApp development, avoiding fallacies in real-world applications and ensuring networks operate efficiently.

\end{abstract}
\section{Introduction}
\label{sec:intro}
The evolution of Open Radio Access Networks (O-RAN){\footnote{A table in Section~\ref{sec:acronyms} lists all acronyms in the paper.}} is accelerating with deployments by multiple operators, such as Vodafone, in worldwide locations~\cite{tim2021oran,baldini2024toward,oran2024map}, including the UK~\cite{vodafone2022oran}.
O-RAN integrates AI-based applications (xApps) to enhance the responsiveness and efficiency of network management~\cite{polese2023understanding}.
Although machine learning models have been used to predict the behaviour of 3G, 4G, and 5G networks~\cite{tahat2024exemplification}, O-RAN's approach to AI integration in 6G significantly advances real-time network adaptability~\cite{chataut20246g}.
{We do not explicitly model such AI-based predictors, but model the probabilistic nature of their output, as accompanied by some level of confidence.
Precisely, we assume that AI-based xApp can learn probabilities for User Equipment (UEs) being on or off, and such probabilities are used to control the network.}

Traditional cellular networks often fail to capture real-world variability~\cite{zaki2015adaptive} and usually use static rule-based policies, such as indefinitely switching off radio cells (RCs) when no users are connected.
These static policies cannot analyse dynamic behaviour, resulting in unclear situations for when and how long before reactivating RCs.
This leads to issues with network availability, compromises in quality of service (QoS), and inefficiencies in power consumption. {  For example, highly frequent RC on/off can result in the deterioration of user-perceived delay, unnecessary frequent handovers of UEs, and higher costs associated with RC mode transitions.\cite{rached2018time,guo2016delay}}
Achieving a balance between availability, QoS, and power efficiency has become a critical challenge for the communications industry.

There have been significant contributions toward addressing these issues in different contexts~\cite{thantharate2024greensky,mohan2024tpemlb,brito2024architecture,singh2024energy,keshta2024game}. 
Recent efforts, such as those in \cite{baldini2024toward,brito2024architecture}, have explored how the O-RAN architecture can support power management and how implementing specific architectural approaches can drastically reduce energy consumption while meeting the requirements for QoS and service availability. Additionally, Zhang et al.\cite{jian2019energy} optimise strategies for switching RCs on and off, as well as user association policies, while ensuring users' QoS is maintained. However, traditional optimisation methods often demand significant computing power and struggle to adapt to the dynamic and evolving nature of network environments.

In contrast, ML-based optimisation methods can overcome such difficulties.
In \cite{yeh2023deep}, an intelligent network application utilising deep learning enables network slicing in O-RAN, allowing emerging IoT services to coexist while ensuring compliance with required service level agreements. Sbella et al. \cite{sesto2019use} employ supervised and unsupervised learning technique to predict the network behaviour and dynamically switch base stations on and off to conserve energy in a mobile cellular network. However, this policy may affect the QoS due to the switching off at the macro layer.
In the context described above, O-RAN and AI-based management applications introduce new challenges that traditional models did not face.
One example is misconfiguration risks~\cite{yungaicela2024misconfiguration}.
Our work aims to bridge this gap by emphasising the role of formal verification in developing xApps for O-RAN. 
This links to a lack of policy that accounts for the necessary dynamics of O-RAN networks. The ML is adopted to predict the dynamics of O-RAN networks based on the data set, and then the formal verification generates the decision based on the prediction.

We strongly advocate for formal verification as an essential part of ensuring that AI-based xApps for O-RAN avoid critical logical fallacies during development.
By analysing dynamics %
early, we can prevent future availability issues caused by static cell-switching policies.
In this paper, we demonstrate, through simulations with a well-known formal verification tool, the PRISM model checker~\cite{kwiatkowska2002prism} that AI-based decisions can improve energy efficiency and availability, providing robust results that can guide real-world development.

In this paper, we outline key contributions that support our position:
\begin{itemize}
  \item We argue that formal verification of xApp models, using tools like PRISM, should be the first step in the development of AI-driven applications for O-RAN.
  \item We present simulations that highlight how AI-based predictions can enhance cell-switching policies.
  By predicting user behaviour, RCs can make informed decisions about when to switch on or off, improving both energy efficiency and service availability.
  \item Our analysis of energy efficiency versus availability thresholds reveals that static policies often lead to self-imposed denial-of-service scenarios.
  Incorporating AI-based predictions mitigates these risks and ensures better resource management.
\end{itemize}

The remainder of this paper is organised as follows. \Cref{sec:bg} provides the necessary background about O-RAN, energy management, and probabilistic model checking for further presentation of our work in the subsequent sections. In \Cref{sec:dynamic}, we introduce a dynamic UE scenario, its corresponding management policy, and discuss considerations in modelling. After that, we present the PRISM model in \Cref{sec:analysis} and discuss the interesting properties and related verification results. Finally, we discuss future work in \Cref{sec:conclusion}.

\section{Background}
\label{sec:bg}

By integrating the controlling of xApps into O-RAN, one can optimise energy efficiency, as demonstrated by two proposed xApps in \cite{liang2024enhancing}: one switches unused RCs to sleep mode, and the other reallocates any UEs while maintaining QoS and balanced RC workload.
We first present  an overview of the structure and components of O-RAN, to link back to current methods for energy management.

\subsection{Overview of O-RAN}
\label{sec:oran}

\begin{figure}[tb]
    \centering
    \includegraphics[width=\textwidth]{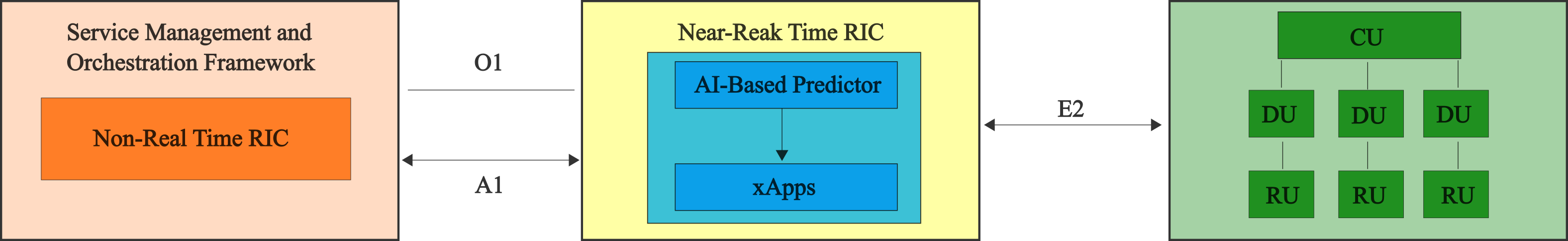}
    \caption{{O-RAN Architecture}}
    \label{fig:O-RAN_Architecture}
\end{figure}

O-RAN adopts and supports the 3\textsuperscript{rd} Generation Partnership Project (3GPP) functional split, in which base station (BS) functions are virtualised as network elements and distributed across various network nodes: the central unit (CU), distributed unit (DU), and radio unit (RU)~\cite{bonati2020open,polese2023understanding}, {as shown in Figure~\ref{fig:O-RAN_Architecture}}.
The CU is typically situated in a central location and handles the higher-layer functions of the RAN. It manages multiple distributed units, enhancing scalability and optimising resource utilisation.
By contrast, the DU is responsible for lower-layer functions such as the radio link control, medium access control, and parts of the physical layer, and is closer to the RU. It handles the real-time processing and efficient management of data transmission between the CU and RU. The RU, located at the network’s edge, manages the radio frequency processing and transmission over the air interface between the BS and UE. 
The core innovation of O-RAN is improving the RAN performance by utilising the above-virtualised network elements and open interfaces, integrating intelligence into the RAN through introducing the RAN Intelligent Controller (RIC)~\cite{bonati2021intelligence}.
This innovation allows operators to implement and deploy custom RAN functions composed of virtual network functions. Besides that, intelligence is required to handle demanding service requirements~\cite{niknam2022intelligent}.

Finally, utilising the RIC in the RAN has the potential to enhance energy efficiency in modern telecommunication systems within the O-RAN architecture.
In \cite{liang2024enhancing}, two xApps are proposed to facilitate RC switching, aiming to reduce power consumption {of static model with fixed location of UEs}.
The first xApp is designed to switch  RCs with a lower number of UEs into sleep mode, while the second xApp reallocates the UEs to some RCs, while ensuring that the QoS for each UE is maintained, and the workload for each RC remains within a specific limit.

\subsection{Formal Methods and PRISM}
\label{sec:fm-prism}
In our context, formal methods refer to mathematically rigorous techniques used in network management to ensure system correctness, reliability, and other properties.
As modern telecommunications systems, such as O-RAN with applications for dynamic control, become increasingly complex, formal methods are essential to mitigate potential risks like misconfigurations or unintended behaviour.

These methods can be applied to validate protocols, optimise resource allocation, and ensure energy-saving strategies do not compromise service availability.
By offering precise mathematical models of system behaviours, formal methods help network engineers analyse edge cases that might be missed by traditional testing techniques.
Formal verification techniques, such as model checking, offer an exhaustive approach to system validation in dynamic environments~\cite{clarke1997model}.

In this work, we model our scenarios with PRISM~\cite{kwiatkowska2011prism}, a leading tool for probabilistic model checking, used to formally verify systems with random or uncertain behaviours.
In network management, where uncertainty prevails (such as fluctuating user demands or varying conditions) probabilistic model checking provides a means to model 
systems as stochastic processes, e.g. Markov chains, and to enable the verification of performance metrics like reliability, availability, and energy efficiency.
In the O-RAN network systems, PRISM is particularly relevant in scenarios where (AI-driven or not) decisions are based on probabilistic models of user behaviour.

\subsection{Probabilistic Model Checking}
Our models take the form of (labelled) continuous-time Markov chains (CTMCs) to represent the real-time behaviours of users and RCs.
A CTMC consists of a finite set of states, with a designated initial state, and a function that assigns a transition \emph{rate} $\lambda$ between any two states.
A transition between states $s$ and $s'$ can occur only if $\lambda(s,s') > 0$, in which case the probability of this transition happening within $t$ time units is $1- e^{-\lambda(s,s')\cdot t}$.
Essentially, this means transitions with a higher rate occur more frequently.
We label states with atomic predicates to simplify the representation of logical formulae over the CTMC, e.g. $s \mapsto \{\text{steady}\}$.

Instead of constructing a CTMC directly, we utilise the PRISM modelling language~\cite{kwiatkowska2002prism}, a state-based language based on Reactive Modules~\cite{alur1999reactive}, that facilitates high-level process specifications. 
Processes are depicted by modules consisting of non-deterministic choices among action-labelled guarded commands that denote transitions.
For example, a module might represent a user with an internal state, e.g., connection, and transitions between internal states at a specified rate modelling the handover.
PRISM has been successfully used in a range of domains including real-time probabilistic communication protocols~\cite{duflot2005probabilistic}, biological systems~\cite{heath2008probabilistic} and human-swarm interactions~\cite{gu2023successful}. It is applicable and expressive enough to perform an exhaustive analysis of all possible behaviours of the system under study~\cite{kwiatkowska2002prism}.

A core component of PRISM is a model checker that allows system behaviours to be quantified~\cite{legay2010statistical}, e.g., querying the probability a system reaching a steady status.
Properties of interest are expressed in an extended form of Continuous Stochastic Logic (CSL)~\cite{baier2003model}, which is a temporal logic with probabilistic operators. For example, the expression  $\mathbf{F}\, \varphi$, where $\mathbf{F}$ is temporal operator meaning \emph{eventually}, asserts that, for all paths, we eventually reach a state in which  $\varphi$ is true.
Quantitative properties are specified with the operator $\mathcal{P}_{=?}$. For instance $\mathcal{P}_{=?}\,[\,\varphi\,]$ represents an expression on the \emph{likelihood} of a path existing  where $\varphi$ is true, rather than \emph{if} a path exists where $\varphi$ is true.
It also possible to express bounded variants, e.g., the property $\mathcal{P}_{=?}\,[\,\mathbf{F}^{\leq t}\,\varphi\,]$ states the likelihood of $\varphi$ to be true within $t$ time units.
The steady-state behaviour of a model is examined by $\mathcal{S}$ operator, similar to $\mathcal{P}$, which can also be applied to quantify the probability, e.g. property $\mathcal{S}_{=?}\,[\,\varphi\,]$ returns the steady-state probability of being in a state that satisfies $\varphi$.
PRISM allows \emph{rewards} (\emph{costs}) to be assigned to states. The $\mathcal{R}$ operator can quantify reward-based properties, e.g. $\mathcal{R}_{=?}\,[\,\mathbf{S}\,]$.In addition, we can also customise properties to obtain results for any single state (not only the initial state) of the model. This is achieved by using \emph{filters}, e.g. $filter\,(\,min,\,\mathcal{P}_{=?}\,[\,\mathbf{F}^{\leq t}\,\varphi\,]\,,\,s_{1}\,)$ gives the minimum value of the probability of reaching $\varphi$ within $t$ time units from state $s_{1}$.

In PRISM, processes are represented by modules consisting of non-deterministic choice over action-labelled guarded commands (which denote transitions); modules are composed of all common actions. For example, a guarded command has the form:
$$[action] \ guard \ \rightarrow rate\ : \ update$$
meaning the process can make a transition to a state described by the $update$ at the given $rate$ if the $guard$ is true. Transitions in different modules can be synchronised with the same action labels. The rate of the synchronisation is then the \emph{product} of individual rates:
$$[action_1] \ guard_1 \ \rightarrow rate_1\ : \ update_1$$
$$[action_1] \ guard_2 \ \rightarrow rate_2\ : \ update_2$$
where the transition to $update_1$ and $update_2$ happens when $guard_1$ and $guard_2$ are true with rate $rate_1 \cdot rate_2$.
\section{Dynamic Client Management}
\label{sec:dynamic}

As discussed in~\Cref{sec:oran}, the two xApps in~\cite{liang2024enhancing} consider only the static scenario where the states of UEs are also unchanged (always on). One important consideration in their xApp2 is Reference Signals Received Power (RSRP), which is a measurement of the signal strength of a UE received from an RC and represents the quality of connection between the UE and the RC. A UE can be re-assigned to another RC only if the new connection satisfies the RSRP threshold. 

\begin{figure}[tb]
    \centering
    \includegraphics[width=0.7\textwidth]{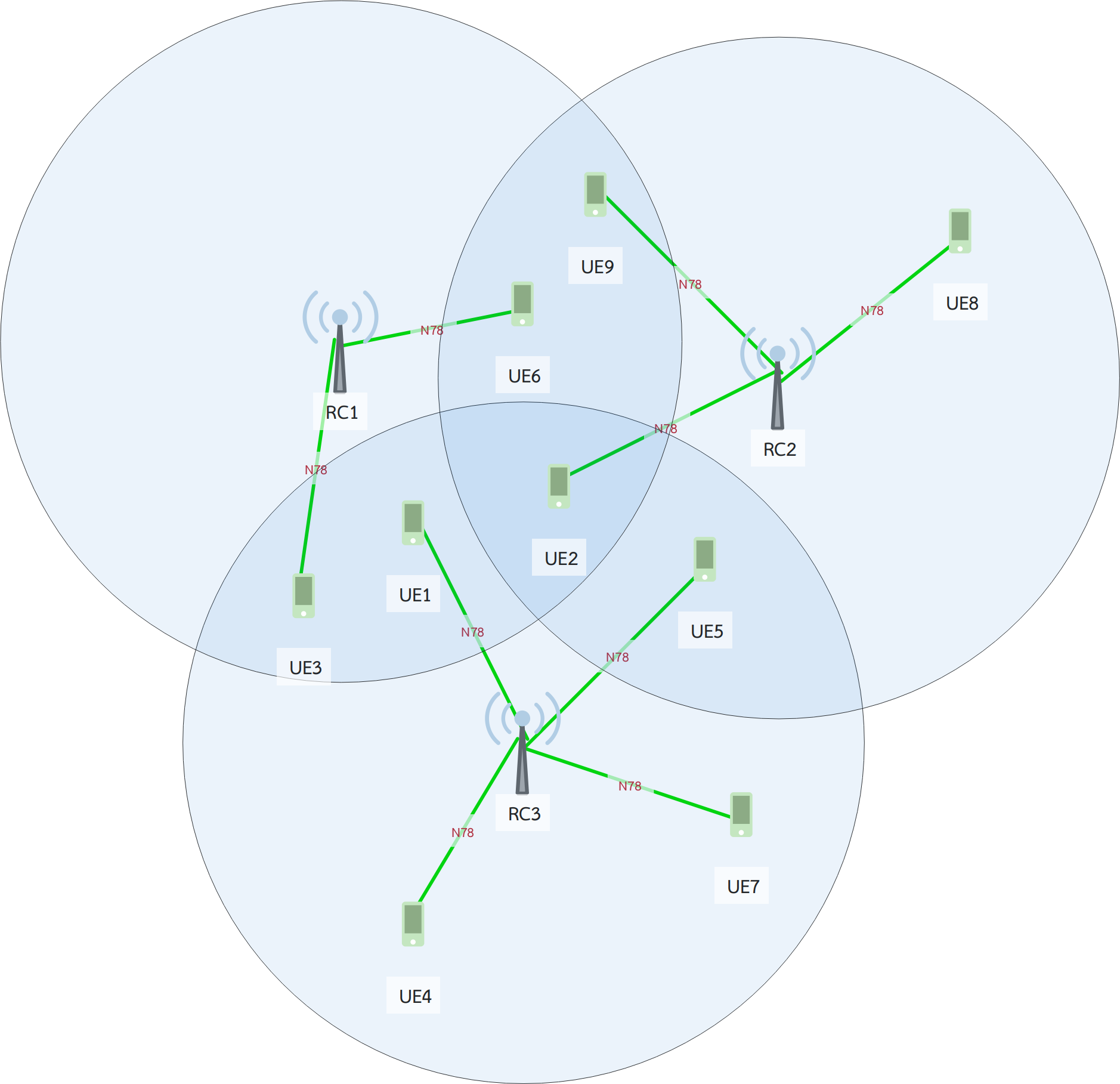}
    \caption{A scenario with 3 RCs and 9 UEs where their locations are all fixed. The diagram shows one possible connection where all the RCs are on.}
    \label{fig:scenarion-9-UEs-3-RCs}
\end{figure}

In this paper, we consider scenarios where UEs are dynamically switched on and off. For example, in a sensor network, sensors are regularly on or off to monitor particular events and save energy, but when a UE is switched on or off is uncertain.
We assume all UEs and RCs's locations are fixed, as illustrated in \Cref{fig:scenarion-9-UEs-3-RCs} where three circles denote the possible coverage areas by the three RCs. 
We further assume the RSRP for a potential connection from each UE to an RC (that covers the UE) satisfies the required minimum threshold so RSRP is not considered in our scenario.  

We design a policy to minimise the total power consumption by RCs while maintaining the QoS for each UE. In this policy, 
\begin{enumerate}[label={\textbf{PO{\arabic*}}}]
    \item each RC has a serving capacity; the RC will not accept a new connection from any UE if it is full in capacity. \label{PO:RC::capacity} 
    \item an RC will be on standby if all its covered UEs are off. \label{PO:RC::off}
    \item when a UE is switched on, \label{PO:UE:connect}
        \begin{enumerate}[label={\textbf{\ref{PO:UE:connect}.{\arabic*}}}, wide, labelwidth=!, labelindent=0pt]
            \item it will randomly choose an RC (that covers it, is on, and not full in capacity) to connect. \label{PO:UE:connect:join}
            \item if no such RC exists, it will randomly choose an RC (that covers it and is on standby) to connect and at the same time, the RC is switched to on from standby. \label{PO:UE:connect:rc_on}
            \item otherwise, the UE fails to connect to any RC so it is out of service. \label{PO:UE:connect:noservice}
        \end{enumerate}
    \item when a UE is switched off, its current connection to the RC is also released. \label{PO:UE::off}
\end{enumerate}
For the scenario considered in \Cref{fig:scenarion-9-UEs-3-RCs}, we expect,  at some point, RC1 to be on standby and the UE3 and UE6 it  currently serves would be served by RC2 and RC3 respectively. As a result, the total power consumption is reduced.

\subsection{Modelling Considerations}
\label{sec:dynamic:modelling}
We aim to model the dynamic UE management using CTMCs in PRISM and then analyse energy consumption and service availability using probabilistic model checking. To model the scenario in \Cref{fig:scenarion-9-UEs-3-RCs}, we consider  
\begin{enumerate}[label={\textbf{C{\arabic*}}}, wide, labelwidth=!, labelindent=0pt]
    \item An RC can be in two states: standby or operating. It also maintains the number of served UEs to ensure its capacity will not be exceeded (otherwise, the service quality for UEs will be deteriorated). \label{C:RC} 
    \item Whether an UE is covered by an RC is recorded in the model but not the exact location of UEs. \label{C:UE:location} 
    \item A UE can be in one of two states: on and off. {But when it will be switched from on to off, or off to on is uncertain, but the probability of leaving on or off follows exponential distributions.} We use $r_\mathit{on}$ (or $r_\mathit{off}$) to denote a UE's on rate (or off rate). The mean duration of on and off for a UE is $1/r_\mathit{on}$ and $1/r_\mathit{off}$.  \label{C:rate} 
    \item An UE tries to connect to an RC based on the policy \ref{PO:UE:connect}. If it succeeds, the connected RC is recorded in the UE and hence is served. At the same time, the number of served UEs in the RC is increased by 1. \label{C:UE:connect} 
    \item The number of served UEs in an RC will be decreased by 1 if a connected UE is switched off. \label{C:UE:off}
    \item Time consideration: each command in a CTMC model of PRISM must have an associated rate. For the commands used to switch on or off an UE, its associated rate is $r_\mathit{on}$ (or $r_\mathit{off}$). Otherwise, each command has a default rate $r$ which is larger than $r_\mathit{on}$ and $r_\mathit{off}$. This  means it has less time or duration than that for the UE's on or Off time. {In this paper, we choose $r$ to be 1.0, and so $r_\mathit{on}$ (or $r_\mathit{off}$) should be larger than it. Another reason to choose 1.0 for $r$ is due to the PRISM's system module which is combined and merged from all the modules by multiplying rates from multiple commands. Simply any rate multiplied by 1 is still itself. Otherwise, the merged rate will be scaled.} \label{C:time}
\end{enumerate}
\section{Formal Model and Analysis}
\label{sec:analysis}

In this section, we explain how we model the problem in PRISM, and how we capture power consumption and availability.  We also present and discuss our results.
We remark that our quantitative results are not limited to our model and offer  important insights for policies that have to balance between power consumption and service availability.

\subsection{PRISM Model Description}
\label{sec:analysis:prism-model}

{
    In PRISM, we model each RC or UE as a module. The behaviour of a RC is defined using a state machine shown in the top diagram in \Cref{fig:prism:model} and that of a UE is defined in the bottom diagram. The connections between these modules are illustrated in the middle diagram in \Cref{fig:prism:model}. A connection between a RC module and a UE module denotes a synchronisation using actions between two modules. There are three possible actions for each connection: such as \texttt{ue9\_to\_rc3} for UE9 to connect to RC3 when RC3 is on, \texttt{ue9\_on\_rc3} for UE9 to connect to RC3 when RC3 is off, and \texttt{ue9\_from\_rc3} for UE9 to disconnect from RC3.
\begin{figure}[tb]
    \centering
    \includegraphics[width=1.0\textwidth]{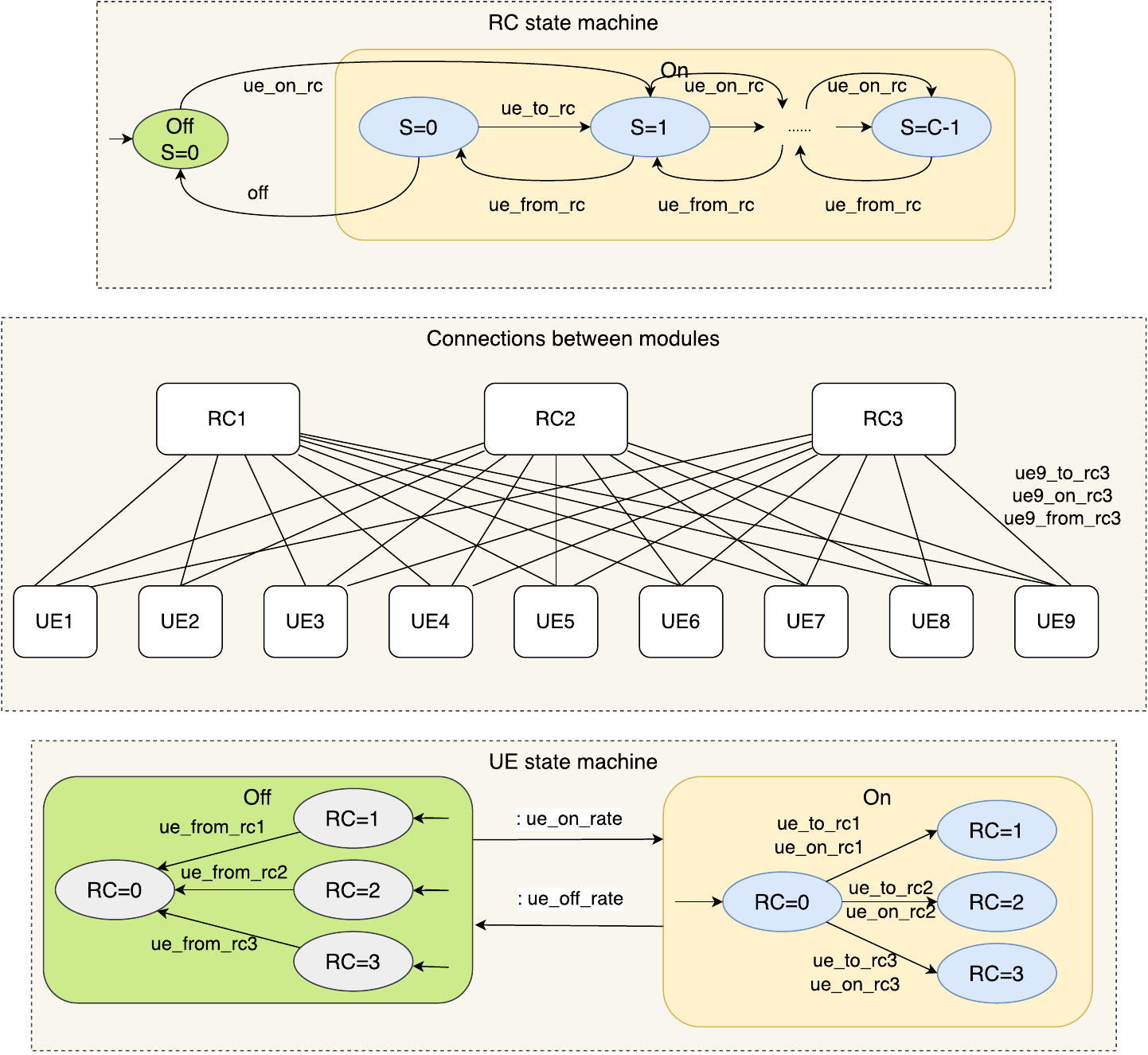}
    \caption{{PRISM module connections and behaviours.}}
    \label{fig:prism:model}
\end{figure}

As given in the RC state machine in the top diagram of \Cref{fig:prism:model}, initially it is \texttt{Off} with serving no UE (\texttt{S=0}) and can be switched on by a demand from any UE through synchronisation over \texttt{ue\_on\_rc}. At the same time, the UE is connected to the RC and the \texttt{S} is increased to 1. Before the RC reaches its capacity (\texttt{S=C-1}), it allows UE to disconnect from it (so \texttt{S} decreases) and UE to connect to it (so \texttt{S} increases). If $S=0$, then it is switched off.

The state machine of a UE, as illustrated in \Cref{fig:prism:model} also has two states: \texttt{Off} and \text{On}. The switch between them is not triggered by actions but uncertainly follows exponential distributions with a rate \texttt{ue\_on\_rate} for \texttt{Off} to \texttt{On} and a rate \texttt{ue\_off\_rate} for \texttt{O1} to \texttt{On}. The UE is initially \texttt{Off} and will be switched on following exponential distributions. Each \texttt{Off} or \texttt{On} contains four substates to record the connected RC id or not connected (\texttt{RC=0}).
}

\begin{figure}[!htb]
    \centering
\begin{lstlisting}[language=PRISM,]
ctmc
const double default_rate = 1; const int N_RCs = 3;
const int rc1_id = 1; const int rc2_id = 2; const int rc3_id = 3;
const int rc1_capacity = 5; const int rc2_capacity = 5; ...
const double power_consumption_rc_on = 1;
const double power_consumption_rc_off = 0.01;
module rc1
  rc1_on : bool init false;  // standby - false, on - true 
  rc1_serving: [0..rc1_capacity] init 0;
  // rc1 serves no UEs, switch off
  [rc1_off] (rc1_serving = 0) & rc1_on -> default_rate: (rc1_on' = false) & (rc1_serving' = 0);
  // A new ue is linked to this rc1
  [ue1_to_rc1] (rc1_serving < rc1_capacity) & rc1_on -> default_rate: (rc1_serving' = rc1_serving + 1);
  [ue2_to_rc1] (rc1_serving < rc1_capacity) & rc1_on -> default_rate: (rc1_serving' = rc1_serving + 1);
  ...
  // Switch on is on demand
  [ue1_on_rc1] (rc1_serving = 0) & (rc1_on = false) -> default_rate: (rc1_on' = true) & (rc1_serving' = rc1_serving + 1);
  [ue2_on_rc1] (rc1_serving = 0) & (rc1_on = false) -> default_rate: (rc1_on' = true) & (rc1_serving' = rc1_serving + 1);
  ...
  // A UE is disconnected from this rc1
  [ue1_from_rc1] (rc1_serving > 0) & rc1_on -> default_rate: (rc1_serving' = rc1_serving - 1);
  [ue2_from_rc1] (rc1_serving > 0) & rc1_on -> default_rate: (rc1_serving' = rc1_serving - 1);
  ...
endmodule
\end{lstlisting}
    \caption{The modelling of an RC in PRISM and the snippet for other two RCs are omitted.}
    \label{fig:model:rc}
\end{figure}

We show the modelling of an RC in PRISM in \Cref{fig:model:rc}.{\footnote{The complete model is available on GitHub at \url{https://github.com/RandallYe/ORAN-xApp-QoS-PRISM-Models/blob/master/DataMod24/ue_dynamics_9_3_cnf1.prism}.}} In the beginning part \mylines{2}{6} of the snippet, we define some constants: the default rate and the number of RCs (set to 1 and 3 respectively \myline{2}), the identity for each RC \myline{3}, the serving capacity for each RC \myline{4}, and the power consumption of an RC in its operating and standby states (\mylinestwo{5}{6}). 

The module \lstinprism{rc1} \mylines{7}{24} contains two local variables: \lstinprism{rc1_on} to record the RC's state and \lstinprism{rc1_serving} to record the number of UEs in serving. The module models the behaviours of RC1. The command \myline{11}  turns the RC to standby and clears the serving record (corresponding to policy \ref{PO:RC::off}). The commands \mylines{13}{15}  connect a UE to this RC and increase \lstinprism{rc1_serving} if it is not full in capacity (corresponding to policy \ref{PO:UE:connect:join}). The commands \mylines{17}{19}  change the RC to the operating state by a UE and increase \lstinprism{rc1_serving} if the RC is in standby (corresponding to policy \ref{PO:UE:connect:rc_on}). The commands \mylines{21}{23}  decrease \lstinprism{rc1_serving} if an UE leaves this RC (corresponding to consideration \ref{C:UE:off}). We note that each command has an associate label such as \lstinprism{rc1_off} for the command \myline{11} and \lstinprism{ue1_to_rc1} for the command \myline{13}. These labels could be used for synchronisation with other modules such as \lstinprism{ue1_to_rc1} used for synchronisation with a module called \lstinprism{rc1} (whose model will be shown later in \Cref{fig:model:ue}), or simply to facilitate the definition of transition rewards or simulation.

\begin{figure}[!htb]
    \centering
\begin{lstlisting}[language=PRISM,]
const double ue1_off_rate = 0.01; 
const double ue1_on_rate = 0.1;
const bool ue1_within_rc1 = true; 
const bool ue1_within_rc2 = false; 
const bool ue1_within_rc3 = true; 
module ue1 
  ue1_on : bool init false; // On and off
  ue1_rc : [0..N_RCs] init 0; // 0 - Disconnected, or connected RC id
  [] ue1_on -> ue1_off_rate: (ue1_on' = false); // On -> Off
  [] (!ue1_on) -> ue1_on_rate: (ue1_on' = true); // Off -> On
  // Connect to one RC if the RC is on and has available capacity
  [ue1_to_rc1] (ue1_on) & (ue1_rc = 0) & ue1_within_rc1 & rc1_on & (rc1_serving < rc1_capacity) -> default_rate: (ue1_rc' = rc1_id); /*
  [ue1_to_rc2] (ue1_on) & (ue1_rc = 0) & ue1_within_rc2 & rc2_on & (rc2_serving < rc2_capacity) -> default_rate: (ue1_rc' = rc2_id);*/
  ...
  // Connect if all RCs in the range are off or full capacity, choose one RC that is within the range and off
  [ue1_on_rc1] (ue1_on) & (ue1_rc = 0) & 
    !(ue1_within_rc1 & rc1_on & (rc1_serving < rc1_capacity)) &
    !(ue1_within_rc2 & rc2_on & (rc2_serving < rc2_capacity)) & 
    !(ue1_within_rc3 & rc3_on & (rc3_serving < rc3_capacity)) & 
    (ue1_within_rc1 & !rc1_on)
    -> default_rate: (ue1_rc' = rc1_id);
  ...
  // Disconnect from the connected rc 
  [ue1_from_rc1] (!ue1_on) & (ue1_rc = rc1_id) -> default_rate: (ue1_rc' = 0);/*
  [ue1_from_rc2] (!ue1_on) & (ue1_rc = rc2_id) -> default_rate: (ue1_rc' = 0);*/
  ...
endmodule
\end{lstlisting}
    \caption{The modelling of a UE in PRISM and the snippet for other UEs are omitted.}
    \label{fig:model:ue}
\end{figure}
The modelling of an UE is illustrated in \Cref{fig:model:ue} where the constants \mylinestwo{1}{2} denote $r_{\textit{on}}$ and $r_{\textit{off}}$ for UE1 (whose values are 0.01 and 0.1, denoting their mean durations are 100 and 10 units of time for the ON and OFF of UE1). The constants \mylines{3}{5} denote whether UE1 can be served by each RC (the values \lstinprism{true}, \lstinprism{false}, and \lstinprism{true} means UE1 can be served by RC1 and RC3, but not by RC2). 

The \lstinprism{ue1} module defined \mylines{6}{25} contains two local variables: \lstinprism{ue1_on} \myline{7} and \lstinprism{uc1_rc} \myline{8} to record the current state and the connected RC of UE1. The two commands \mylinestwo{9}{10} switch on and off UE1 with the corresponding rates (see \ref{C:rate}). The command \myline{12} allows UE1 to connect to RC1 (\lstinprism{ue1_rc' = rc1_id}) if the UE is ON (\lstinprism{ue1_on}) and has not been connected (\lstinprism{ue1_rc = 0}), UE1 can be served by RC1 (\lstinprism{ue1_within_rc1}), RC1 is operating (\lstinprism{rc1_on}) and has the available capacity (\lstinprism{rc1_serving < rc1_capacity}) to serve UE1. This command will synchronise (because they have the same label \lstinprism{ue1_to_rc1}) with the command \myline{13} in \Cref{fig:model:rc} to let UE1 connect to RC1. This command implements policy \ref{PO:UE:connect:join} and consideration \ref{C:UE:connect}. There are other two similar commands for UE2 and UE3, omitted here for simplicity. If both RC1 and RC3 are available to serve UE1, which UE1 will be connected is randomly chosen (according to the PRISM's semantics). 

The policy \ref{PO:UE:connect:rc_on} and consideration \ref{C:UE:connect} is realised through the synchronisation of the command \mylines{15}{20} with the command \myline{17} in \Cref{fig:model:rc}. This is to deal with a situation when all RCs are either on standby, out of the range, or full in capacity. The command will let UE1 to connect to RC1 if RC1 is on standby and within the range of UE1. Similarly, there are other two similar commands for UE2 and UE3, omitted here for simplicity. If both RC1 and RC3 are in standby, then which UE1 will be connected is randomly chosen.  

The command \myline{23} allows UE1 to be disconnected from RC1 if UE1 is currently connected to RC1 and is off now. Similarly, there are other two similar commands for RC2 and RC2.

\subsection{Verification and Evaluation}
\label{sec:analysis:verification}
Once the PRISM model is constructed, we can use probabilistic model checking to query interesting  qualitative and quantitative properties. 

The first property which we are interested in is the average total power consumption of all RCs. We add the following two rewards to the PRISM model which assign \lstinprism{power_consumption_rc_off} and \lstinprism{power_consumption_rc_on} to each RC if it is standby and operating.

\begin{lstlisting}[language=PRISM,]
rewards "standby"
  !rc1_on : power_consumption_rc_off;
  !rc2_on : power_consumption_rc_off;
  !rc3_on : power_consumption_rc_off;
endrewards
rewards "operating"
  rc1_on : power_consumption_rc_on;
  rc2_on : power_consumption_rc_on;
  rc3_on : power_consumption_rc_on;
endrewards
\end{lstlisting}

\begin{table}[!htb]
    \caption{Interested properties and their CSL formulas.}
\label{tab:props}\centering
\begin{tabular}{@{}ll@{}}
    \toprule
    Property & CSL formula \\
    \midrule
    P1 & \lstinprism{R\{"standby"\}=?[S]} \\
    P2 & \lstinprism{R\{"operating"\}=?[S]} \\
    P3.1 & \lstinprism{S=? [rc1_on]} \\
    P3.2 & \lstinprism{S=? [rc2_on]} \\
    P3.3 & \lstinprism{S=? [rc3_on]} \\
    P4.1 & \lstinprism{filter(min, P=? [F<=t ue1_rc>0], ue1_on&ue1_rc=0)} \\
    P4.2 & \lstinprism{filter(min, P=? [F<=t ue2_rc>0], ue2_on&ue2_rc=0)} \\
    \dots & \dots \\ 
    P4.9 & \lstinprism{filter(min, P=? [F<=t ue9_rc>0], ue9_on&ue9_rc=0)} \\
    \bottomrule
\end{tabular}
\end{table}

Then we can use CSL formulas \lstinprism{R\{"standby"\}=?[S]} or \lstinprism{R\{"operating"\}=?[S]} to get the long-run average power consumption per one unit of time, as shown in \Cref{tab:props} and denoted as P1 and P2. The total power consumption is just their sum: P1 + P2. P1 (or P2) provides us with the total power consumption for all three RCs if they are on standby (or operating). We may also want to know the percentage each RC contributes to P1 and P2. This can be queried using the formula \lstinprism{S=? [rc1_on]} which means the long-run probability of being in a state when RC1 is operating. Similarly, we can query that for RC2 and RC3. We use P3.1, P3.2, and P3.3 to denote the property for each RC. 

In addition to power consumption, the service availability for each UE is also an important measurement for a policy. We expect every UE can be connected to an RC within an amount of time. A formula \lstinprism{filter(min, P=? [F<=t ue1_rc>0], ue1_on & ue1_rc=0)} gives the \lstinprism{min}imum probability (\lstinprism{P=?}) of UE1 finally being connected to any RC (\lstinprism{ue1_rc>0}) within $t$ units of time (\lstinprism{F<=t}) when it is switched on but not connected (\lstinprism{ue1_on&ue1_rc=0)}). {This represents the lower-bound guarantee of the service.} A \lstinprism{filter} in PRISM allows us to quantify properties from any state (not only from the initial states of the model usually). In P4.1, the time 0 starts from the states in which UE1 is on but not yet connected. {P4.1, P4.2, or \dots P4.9 are properties for different UEs in a same configuration.}

\paragraph{Configurations.} We consider six configurations of constants shown in \Cref{tab:conf:base}. In this, Cnf1 is a base configuration and other configurations have only one change based on Cnf1. Cnf2 changes the RC capacity from 5 to 6. Cnf3 changes the RC capacity from 5 to 4. Cnf4 increases the rate $r_{off}$ from 0.1 to 1 (so UEs are mostly ON and will only be switched off very quickly). Cnf5 swaps the ON and OFF rates. Cnf6 considers UE3 only being in RC1.

\begin{table}[!htb]
    \caption{Constant values for six configurations where $r$ is for the default rate, Power on and off for \lstinprism{power_consumption_rc_on} and \lstinprism{power_consumption_rc_off}, and Location for the relative position of UEs.}
\label{tab:conf:base}\centering
\small
\begin{tabularx}{\textwidth}{@{}lccccccc@{}}
    \toprule
    Id  & $r$  & Capacity  & Power on  & Power off  & $r_{\textit{on}}$  & $r_{\textit{off}}$  & Location \\
    \midrule
    Cnf1 & 1 & 5 & 1 & 0.01 & 0.01 & 0.1 & as \Cref{fig:scenarion-9-UEs-3-RCs} \\
    Cnf2 & 1 & \textcolor{red}{6} & 1 & 0.01 & 0.01 & 0.1 & as \Cref{fig:scenarion-9-UEs-3-RCs} \\
    Cnf3 & 1 & \textcolor{red}{4} & 1 & 0.01 & 0.01 & 0.1 & as \Cref{fig:scenarion-9-UEs-3-RCs} \\
    Cnf4 & 1 & 5 & 1 & 0.01 & 0.01 & \textcolor{red}{1} & as \Cref{fig:scenarion-9-UEs-3-RCs} \\
    Cnf5 & 1 & 5 & 1 & 0.01 & \textcolor{red}{0.1} & \textcolor{red}{0.01} & as \Cref{fig:scenarion-9-UEs-3-RCs} \\
    Cnf6 & 1 & 5 & 1 & 0.01 & 0.01 & 0.1 & \textcolor{red}{UE3 to be only in RC1} \\
    \bottomrule
\end{tabularx}
\end{table}

\paragraph{Power consumption.}  Model checking shows that the value (0.007346391) of P1 is negligible compared to that (2.265360891) of P2 in Cnf1. So here we only discuss P2 and P3. \Cref{fig:results:power} shows the analysis results of P3 for each RC in the six configurations. The sum of three sections for P3.1, P3.2, and P3.3 is the total height of the bar for a configuration and is also the total power consumption (P2) of RCs because \lstinprism{power_consumption_rc_on} is set to 1 in all configurations. From the diagram, we conclude that Cnf6 has the highest energy consumption, followed by Cnf3. We also note that Cnf5 uses the least energy. This is because Cnf6 has UE3, UE4, UE7, and UE8 only within the range of RC1, RC3, and RC2. Therefore, all the three RCs are operating nearly all the time. Cnf3 has the lowest capacity (4) for each RC.
To serve all UEs, similarly, all the RCs are operating most of the time. Cnf5 has UEs mostly OFF (its mean duration is 100) compared to the mean duration (10) of ON. So Cnf5 has the lowest power consumption because RCs can be in standby when all UEs are off. We also observe that the results of P3.2 and P3.3 are always very close, but that of P3.1 has changed dramatically.
This is because both RC2 and RC3 in \Cref{fig:scenarion-9-UEs-3-RCs} have at least one UE (UE8, or UE4 and UE7) that is within their ranges but cannot be reached by other RCs. In Cnf2, P3.1 is very small and so RC1 is almost not operating, thanks to the high capacity (6) in Cnf2 (so RC2 and RC3 always can serve all UEs).

\begin{figure}[!htb]
    \centering
\pgfplotstableread{
Label P3.1 P3.2 P3.3 topper
Cnf1 0.269176187 0.996428391 0.999756252 0.001
Cnf2 0.01148031 0.999434069 0.999968971 0.001
Cnf3 0.991838656 0.999105787 0.912019495   0.001
Cnf4 0.10443116 0.999943207 0.999999555  0.001
Cnf5 0.179333302 0.240934078 0.327372283   0.001
Cnf6 0.987874498 0.987874498 0.998495534  0.001
}\testdata

\begin{tikzpicture}[thick,scale=0.8, every node/.style={scale=0.8}]
    \begin{axis}[
        ybar stacked,
        ymin=0,
        ymax=3.5,
        xtick=data,
        legend style={cells={anchor=west}, legend pos=north west},
        reverse legend=false, %
        xticklabels from table={\testdata}{Label},
        xticklabel style={text width=2cm,align=center},
    ]
    \addplot [fill=green!80] table [y=P3.1, meta=Label, x expr=\coordindex] {\testdata};
    \addlegendentry{P3.1}
    \addplot [fill=blue!60] table [y=P3.2, meta=Label, x expr=\coordindex] {\testdata};
    \addlegendentry{P3.2}
    \addplot [fill=red!60] table [y=P3.3, meta=Label, x expr=\coordindex] {\testdata};
    \addlegendentry{P3.3}
    \addplot [nodes near coords,point meta=y,nodes near coords style={anchor=south}] table [y=topper, meta=Label, x expr=\coordindex] {\testdata};

    \end{axis}
    \end{tikzpicture}
    \caption{Comparison of power consumption of each RC for six configurations.}
    \label{fig:results:power}
\end{figure}
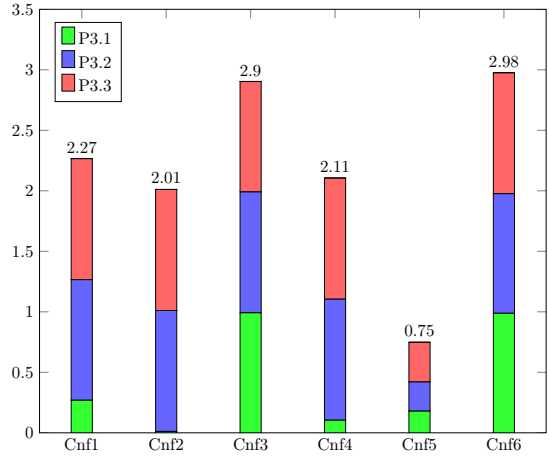

\begin{figure}[!htb]
    \begin{center}
\begin{tikzpicture}[thick,scale=0.8, every node/.style={scale=0.8}]
\begin{axis}[%
    xlabel={$t$},
    ylabel={Minimum Probability},
    y tick label style={/pgf/number format/.cd,fixed,fixed zerofill,precision=2},
    grid=both,
    grid style={line width=.1pt, draw=gray!30},
]
    \addplot table[header=false,col sep=&,row sep=\\,y expr={\thisrowno{1}}] {
1 & 0.629586558\\
2 & 0.859263337\\
3 & 0.94338669\\
4 & 0.97450191\\
5 & 0.98628376\\
6 & 0.990988293\\
7 & 0.993078591\\
8 & 0.994182119\\
9 & 0.994894337\\
10 & 0.995434927\\
11 & 0.99588657\\
12 & 0.996281843\\
13 & 0.996634858\\
14 & 0.996952798\\
15 & 0.997240132\\
    } node[below,pos=0.8] {ue1-3,5,6,8,9};

    \addplot table[header=false,col sep=&,row sep=\\,y expr={\thisrowno{1}}] {
1 & 0.004977367\\
2 & 0.02489165\\
3 & 0.055115364\\
4 & 0.089716702\\
5 & 0.125330242\\
6 & 0.160428756\\
7 & 0.194411199\\
8 & 0.227083342\\
9 & 0.258409644\\
10 & 0.288422773\\
11 & 0.317171961\\
12 & 0.344721188\\
13 & 0.371127582\\
14 & 0.396444825\\
15 & 0.420724649\\
    } node[above,pos=0.7] {ue4,7};

  \end{axis}
\end{tikzpicture}
    \end{center}
    \caption{Comparison of P4 for each UE in Cnf1 where the lower line corresponds to (overlapped) UE4 and UE7 and the upper line corresponds to other (overlapped) UEs.}
    \label{fig:results:cnf1:P4}
\end{figure}
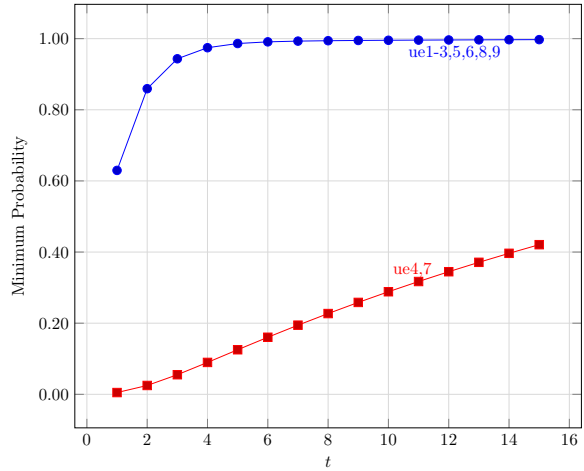

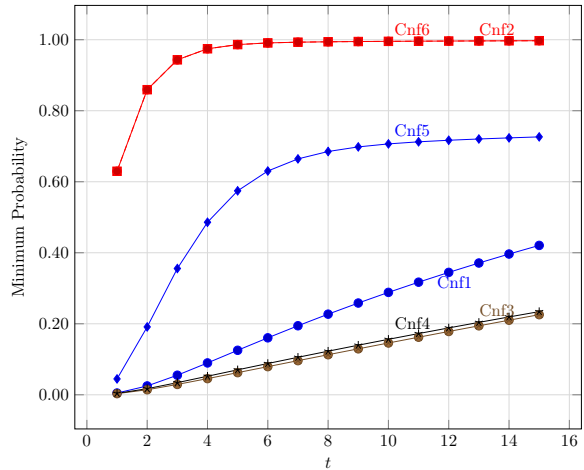
\begin{figure}[!htb]
    \begin{center}
\begin{tikzpicture}[thick,scale=0.8, every node/.style={scale=0.8}]
\begin{axis}[%
    xlabel={$t$},
    ylabel={Minimum Probability},
    y tick label style={/pgf/number format/.cd,fixed,fixed zerofill,precision=2},
    grid=both,
    grid style={line width=.1pt, draw=gray!30},
]
    \addplot table[header=false,col sep=&,row sep=\\,y expr={\thisrowno{1}}] {
1 & 0.004977367\\
2 & 0.02489165\\
3 & 0.055115364\\
4 & 0.089716702\\
5 & 0.125330242\\
6 & 0.160428756\\
7 & 0.194411199\\
8 & 0.227083342\\
9 & 0.258409644\\
10 & 0.288422773\\
11 & 0.317171961\\
12 & 0.344721188\\
13 & 0.371127582\\
14 & 0.396444825\\
15 & 0.420724649\\
    } node[below,pos=0.8] {Cnf1};

    \addplot table[header=false,col sep=&,row sep=\\,y expr={\thisrowno{1}}] {
1 & 0.629586558\\
2 & 0.859263337\\
3 & 0.94338669\\
4 & 0.97450191\\
5 & 0.986283759\\
6 & 0.990988292\\
7 & 0.99307859\\
8 & 0.994182119\\
9 & 0.994894337\\
10 & 0.995434927\\
11 & 0.99588657\\
12 & 0.996281843\\
13 & 0.996634858\\
14 & 0.996952798\\
15 & 0.997240132\\
    } node[above,pos=0.9] {Cnf2};

    \addplot table[header=false,col sep=&,row sep=\\,y expr={\thisrowno{1}}] {
1 & 0.00324026\\
2 & 0.014225235\\
3 & 0.029183337\\
4 & 0.045573058\\
5 & 0.062394153\\
6 & 0.07928762\\
7 & 0.096123393\\
8 & 0.11284943\\
9 & 0.129440653\\
10 & 0.145883401\\
11 & 0.162171611\\
12 & 0.178298618\\
13 & 0.194259202\\
14 & 0.210041664\\
15 & 0.225638714\\
    } node[above,pos=0.9] {Cnf3};

    \addplot table[header=false,col sep=&,row sep=\\,y expr={\thisrowno{1}}] {
1 & 0.00402427\\
2 & 0.017162492\\
3 & 0.034229152\\
4 & 0.052250602\\
5 & 0.070297759\\
6 & 0.088110867\\
7 & 0.105618622\\
8 & 0.122802981\\
9 & 0.13966194\\
10 & 0.156198733\\
11 & 0.172418398\\
12 & 0.188326577\\
13 & 0.203929079\\
14 & 0.219231706\\
15 & 0.234240196\\
    } node[above,pos=0.7] {Cnf4};

    \addplot table[header=false,col sep=&,row sep=\\,y expr={\thisrowno{1}}] {
1 & 0.044683691\\
2 & 0.19114485\\
3 & 0.355655439\\
4 & 0.485938543\\
5 & 0.574419249\\
6 & 0.630242721\\
7 & 0.664377454\\
8 & 0.685191658\\
9 & 0.698156802\\
10 & 0.70659928\\
11 & 0.712465175\\
12 & 0.716869166\\
13 & 0.720442277\\
14 & 0.723539507\\
15 & 0.72635999\\
    } node[above,pos=0.7] {Cnf5};

    \addplot table[header=false,col sep=&,row sep=\\,y expr={\thisrowno{1}}] {
1 & 0.629586558\\
2 & 0.859263337\\
3 & 0.94338669\\
4 & 0.974501909\\
5 & 0.986283759\\
6 & 0.990988292\\
7 & 0.99307859\\
8 & 0.994182117\\
9 & 0.994894335\\
10 & 0.995434925\\
11 & 0.995886568\\
12 & 0.99628184\\
13 & 0.996634854\\
14 & 0.996952794\\
15 & 0.997240127\\
    } node[above,pos=0.7] {Cnf6};

  \end{axis}
\end{tikzpicture}
    \end{center}
    \caption{Comparison of P4 for UE4 in different configurations.}
    \label{fig:results:P4:ue4}
\end{figure}

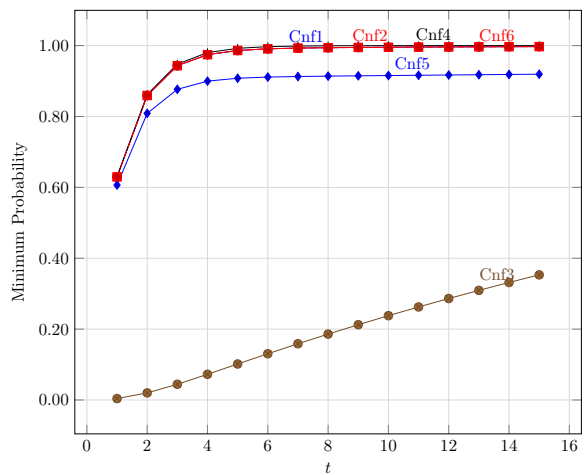
\begin{figure}[!htb]
    \begin{center}
\begin{tikzpicture}[thick,scale=0.8, every node/.style={scale=0.8}]
\begin{axis}[%
    xlabel={$t$},
    ylabel={Minimum Probability},
    y tick label style={/pgf/number format/.cd,fixed,fixed zerofill,precision=2},
    grid=both,
    grid style={line width=.1pt, draw=gray!30},
]
    \addplot table[header=false,col sep=&,row sep=\\,y expr={\thisrowno{1}}] {
1 & 0.629586558\\
2 & 0.859263337\\
3 & 0.94338669\\
4 & 0.97450191\\
5 & 0.986283759\\
6 & 0.990988292\\
7 & 0.99307859\\
8 & 0.994182119\\
9 & 0.994894337\\
10 & 0.995434927\\
11 & 0.99588657\\
12 & 0.996281843\\
13 & 0.996634858\\
14 & 0.996952798\\
15 & 0.997240132\\
    } node[above,pos=0.45] {Cnf1};

    \addplot table[header=false,col sep=&,row sep=\\,y expr={\thisrowno{1}}] {
1 & 0.629586558\\
2 & 0.859263337\\
3 & 0.94338669\\
4 & 0.97450191\\
5 & 0.986283759\\
6 & 0.990988292\\
7 & 0.99307859\\
8 & 0.994182119\\
9 & 0.994894337\\
10 & 0.995434927\\
11 & 0.99588657\\
12 & 0.996281843\\
13 & 0.996634858\\
14 & 0.996952798\\
15 & 0.997240132\\
    } node[above,pos=0.6] {Cnf2};

    \addplot table[header=false,col sep=&,row sep=\\,y expr={\thisrowno{1}}] {
1 & 0.003986534\\
2 & 0.019986222\\
3 & 0.044385164\\
4 & 0.072478658\\
5 & 0.101581831\\
6 & 0.130468612\\
7 & 0.158649677\\
8 & 0.185958967\\
9 & 0.212360833\\
10 & 0.23786847\\
11 & 0.26251147\\
12 & 0.286323548\\
13 & 0.309338179\\
14 & 0.331587217\\
15 & 0.353100592\\
    } node[above,pos=0.9] {Cnf3};

    \addplot table[header=false,col sep=&,row sep=\\,y expr={\thisrowno{1}}] {
1 & 0.63028575\\
2 & 0.861967003\\
3 & 0.947978098\\
4 & 0.980219123\\
5 & 0.992416329\\
6 & 0.997070637\\
7 & 0.998860812\\
8 & 0.999554331\\
9 & 0.999824732\\
10 & 0.999930759\\
11 & 0.999972539\\
12 & 0.999989073\\
13 & 0.99999564\\
14 & 0.999998256\\
15 & 0.999999301\\
    } node[above,pos=0.75] {Cnf4};

    \addplot table[header=false,col sep=&,row sep=\\,y expr={\thisrowno{1}}] {
1 & 0.606579662\\
2 & 0.808857748\\
3 & 0.876673862\\
4 & 0.899767553\\
5 & 0.907983197\\
6 & 0.911246064\\
7 & 0.912857009\\
8 & 0.913913824\\
9 & 0.914781644\\
10 & 0.915581942\\
11 & 0.916355156\\
12 & 0.917114779\\
13 & 0.917865339\\
14 & 0.918608381\\
15 & 0.919344462\\
    } node[above,pos=0.7] {Cnf5};

    \addplot table[header=false,col sep=&,row sep=\\,y expr={\thisrowno{1}}] {
1 & 0.629586558\\
2 & 0.859263337\\
3 & 0.94338669\\
4 & 0.97450191\\
5 & 0.986283759\\
6 & 0.990988292\\
7 & 0.99307859\\
8 & 0.994182119\\
9 & 0.994894337\\
10 & 0.995434927\\
11 & 0.99588657\\
12 & 0.996281843\\
13 & 0.996634858\\
14 & 0.996952798\\
15 & 0.997240132\\
    } node[above,pos=0.9] {Cnf6};

  \end{axis}
\end{tikzpicture}
    \end{center}
    \caption{Comparison of P4 for UE8 in different configurations.}
    \label{fig:results:P4:ue8}
\end{figure}

\paragraph{Service availability.}
We show the results of P4 for each UE in \Cref{fig:results:cnf1:P4} when using Cnf1. For UE4 and UE7, the minimum probability is increased linearly in terms of time (t). It can reach nearly 0.5 after 20 units of time. Their probabilities are relatively small compared to that of other UEs. For other UEs, their minimum probability rises sharply and can reach 0.97 after 4 units of time. This reflects UE4 and UE7 are not well served (because they can only be served by RC3) but others are. This is due to the capacity of RC3 is 5 while it covers 6 UEs. At some points, UE4 or UE7 cannot be served by RC3 because it might be full in capacity when UE4 or UE7 tries to connect.

In \Cref{fig:results:P4:ue4}, we show the results of P4 for UE4 (which is only within the range of RC3) in six different configurations. In Cnf2 and Cnf6 (the tallest two overlapped lines), UE4 is well served and highly likely it can connect to a RC within 4 units of time because RC2 has enough capability to allow all UEs within its range (including \lstinprism{ue4}) to join the network. P4 of UE4 for other configurations are less well served. P4 in Cnf3 and Cnf4 has the least minimum probabilities because RCs have a limited capacity 4 in Cnf3 and UEs are mostly on in Cnf4. 

Similarly, the P4 for UE8 (which is only within the range of RC2) in different configurations is shown in~\Cref{fig:results:P4:ue8}. The UE8 is well served in Cnf1, Cnf2, Cnf4, and Cnf6. It is not well connected in Cnf3 because of the lower capacity (4). 

\subsection{Discussions}
From the analysis of results in terms of power consumption and service availability, we conclude that the service capacity of each RC is a very important parameter to consider. Its capacity should be larger than the number of all UEs within its range. The location of UEs and RCs is also equally important. If each UE can be served by more than one RC, then some RCs could be in standby to save energy.

{
It is noteworthy that our preliminary experiments yield quantitative results in the long term (steady-state analysis) which can be leveraged to strike a balance between quality of service and power consumption.
In themselves, these small-scale models do not achieve optimal solutions (not suitable for use as controlling xApps in O-RAN).
To arrive to optimal solutions, one must first consider the specific application context, taking into account factors such as priority assignment and minimum bounds for both quality of service and power consumption. Additionally, other relevant metrics should be incorporated to capture the nuances of the scenario under investigation.
At present, our models are only capable of providing a minimum guarantee for the network to ensure either service level is met.
}

{
In terms of scalability, our experiments showed that our models can suffer the state explosion when the number of users and RCs increases, resulting in a large computation effort. In this case, exhaustive model checking can be costly. To address this problem, PRISM’s built-in discrete-event simulator enables the use of \emph{statistical model checking} (SMC)~\cite{younes2002probabilistic}, providing approximately accurate results. It effectively samples the model space through repeated simulation instead of exhaustive search~\cite{butkova2019modest,hensel2021probabilistic}. Previous studies have shown that SMC can achieve high performance with large-scale models without a huge cost of accuracy~\cite{gu2023successful},  which is important for practical implementations, considering the requirements of fast response time in O-RAN. However, SMC naturally is not suitable for long-term verification, such as the steady-state analysis, in which case other techniques can be introduced to tackle the state explosion, such as the bonded model checking~\cite{clarke2011model}. 
} %
\section{Conclusion}
\label{sec:conclusion}

This work introduces probabilistic model checking, prior to the development of xApps in O-RAN to quantitatively evaluate QoS like energy efficiency and service availability in a given policy. We have based our work on the two xApps discussed in \cite{liang2024enhancing} where only static scenarios are considered. In this work, we  introduced a dynamic scenario where UEs are regularly (but uncertainly) switched on and off. This requires a new policy to achieve energy efficiency while maintaining service availability. We  modelled the scenario and the policy in PRISM and used probabilistic model checking to explore the QoS in terms of different rates, capacities, and locations of UEs and RCs. The analysis results have shown that RC capacity and redundancy are very important design parameters.

As previously discussed, RSRP is an important measurement to maintain service quality for UEs. We have omitted it and assume UEs are static in this work. We will consider a new dynamic scenario where UEs are moving and RSRP is changing too. Though it is not possible to model the exact location of UEs and its dynamics in PRISM, we could model discretised space and verify the impact of a UE's movement on whether the UE's RSRP becomes over and below a threshold. {PRISM has been applied to analyse the movement of a robot in two dimensions~\cite{Younes2006}, and Signal Spatio-Temporal Logic (SSTL)~\cite{Nenzi2018} can be used to specify spatio-temporal properties for models with linear time and discrete space and verified using SMC.} 

Our dynamic UE management policy and the analysis results could be helpful to implement a similar scenario in O-RAN simulation test-bed and to evaluate the effectiveness of our work in the real O-RAN network.

While O-RAN xApps are evolving to use machine learning for dealing with the complexity of the network environment. To this end, we will also explore the development of methods for learning and adapting policies at run-time and in response to changes in the network environment, e.g., as in \cite{Alrajeh2021}, to avoid fallacies and ensure efficient network operation in O-RAN.
{Future research directions may involve obtaining formally verified outcomes for simplified network architectures, which can then be used to continuously train intelligent xApps and enhance their predictive capabilities.}

{
\section*{Acronyms}
\label{sec:acronyms}

\begin{tabular}{ll}
\textbf{3GPP} & the 3\textsuperscript{rd} Generation Partnership Project
\\\noindent
\textbf{AI} & Artificial Intelligence 
\\\noindent
\textbf{BS} & Base Station 
\\\noindent
\textbf{CSL} & Continuous Stochastic Logic  
\\\noindent
\textbf{CTMC} & Continuous-Time Markov Chain 
\\\noindent
\textbf{CU} & Central Unit 
\\\noindent
\textbf{DU} & Distributed Unit 
\\\noindent
\textbf{IoT} & Internet of Things 
\\\noindent
\textbf{O-RAN} & Open Radio Access Network
\\\noindent
\textbf{QoS} & Quality of Service 
\\\noindent
\textbf{RSRP} & Reference Signals Received Power 
\\\noindent
\textbf{RU} & Radio Unit 
\\\noindent
\textbf{RC} & Radio Cell
\\\noindent
\textbf{RIC} & RAN Intelligent Controller  
\\\noindent
\textbf{SMC} & Statistical Model Checking 
\\\noindent
\textbf{UE} & User Equipment
\\\noindent
\end{tabular}
}
\subsubsection*{\ackname} 
The EPSRC and DSIT support this work through the Communications Hub for Empowering Distributed Cloud Computing Applications and Research (CHEDDAR) under grants EP/X040518/1, EP/Y037421/1 and EP/Y019229/1.

\clearpage

\bibliographystyle{splncs04}
\bibliography{refz}

\end{document}